\documentclass[useAMS,usenatbib]{mnras}
\usepackage{natbib}
\usepackage{aas_macros}
\usepackage{graphicx}
\usepackage{savesym}
\usepackage{listings}
\usepackage[intlimits]{amsmath}
\savesymbol{iint}
\usepackage{txfonts}
\usepackage{bm}
\restoresymbol{TXF}{iint}
\usepackage{amssymb}
\usepackage{hyperref}
\usepackage{amsmath}
\usepackage{tabularx} 
\usepackage{color}
\usepackage{soul}
\usepackage{ulem}
\newcommand{\cE}{{\cal E}}

\newcommand{\cB}{{\cal B}}
\newcommand{\vv}[1]{\bmath{#1} }

\newcommand{\etext}[1]{\quad\mbox{#1}\quad}
\newcommand{\spr}[2]{(\bmath{#1} \cdot \bmath{#2})}
\newcommand{\vpr}[2]{\bmath{#1} \!\times\! \bmath{#2}}
\newcommand{\vdiv}{\bmath\nabla\!\cdot\!}

\newcommand{\vcurl}[1]{\vpr{\nabla}{#1}}

\newcommand{\oder}[2]{\frac{d #1}{d #2}}

\newcommand{\pder}[2]{\frac{\partial #1}{\partial #2}}

\newcommand{\ort}[1]{ \bmath{e}_{#1} }

\newcommand{\beq}{\begin{equation}}
\newcommand{\eeq}{\end{equation}}

\def\A{Alfv\'en }

\def\wb{\bar{b}}
\def\wB{\bar{B}}
\def\wE{\bar{E}}

\def\apjl{Astrophys. J. Lett.}
\def\apjs{Astrophys. J. Supp. Ser. }

\def\aap{Astron. Astrophys. }

\title[Polarisation of  relativistic Alfv\'en waves]{On the polarisation of large-amplitude relativistic Alfv\'en waves in magnetohydrodynamics and force-free degenerate electrodynamics}
\author[S. S. Komissarov] 
{
Serguei S. Komissarov \thanks{E-mail: s.s.komissarov@leeds.ac.uk}\\
School of Mathematics, The University of Leeds, Leeds, LS2 9JT, UK}

\begin{document}
\date{Received/Accepted}
\maketitle

\begin{abstract} 
It is known that in relativistic magnetohydrodynamics, \A waves  are generally elliptically polarised and that in some special inertial frames their polarisation reduces to circular. In contrast, the polarisation of \A waves in magnetodynamics (force-free degenerate electrodynamics), which can be interpreted as magnetohydrodynamics in the limit of infinitely high plasma magnetisation, \A waves have linear polarisation. To resolve this apparent conflict, the properties of relativistic \A waves are re-examined, paying particular attention to the rest frame of undisturbed plasma through which these waves propagate. It is shown that in this frame, the polarisation of \A waves is always elliptical, reducing to circular only in the Newtonian limit. As plasma magnetisation increases, the polarisation ellipse becomes highly elongated, with its eccentricity approaching unity, and its major axis growing in size without limit. This explains why \A waves are linearly polarised in magnetodynamics. The inertial frames where magnetohydrodynamic \A waves have circular polarisation are not comoving with the undisturbed plasma, and in the limit of infinite magnetisation, either their speed reaches the speed of light, or the \A wave reduces to pure electromagnetic wave.           
\end{abstract}   
                                                                                       

\section{Introduction}
\label{introduction}

\A waves are fundamental waves of magnetohydrodynamics. They are very important in the physics of the solar corona and wind, providing a  mechanism of effective energy transport to large distances where the observations indicate in situ heating of plasma \citep{GBML94,GSB95,BC13review}. They have also been considered as drivers of stellar winds from other types of stars \citep[e.g.][]{BO-75,HFL-83,YSK-19}.

\A waves are also expected to play an important role in magnetospheres, jets, and winds of relativistic objects, such as neutron stars and black holes.  For example, the non-linear interactions of \A waves driving plasma turbulence and its heating were analysed in \citep{TB-98}. \A waves launched by star quakes of highly-magnetised neutron stars (magnetars) have been related to gamma-ray and x-ray bursts \citep[GRBs,][]{BBGM-89, DT-92}, as well as fast radio bursts \citep[FRBs, see][and references therein]{Zhang23}. \A waves may be the key component of the coherent radio emission mechanism in FRBs \citep{LyutikovCE-21}.  In pulsars, they may cause glitches of pulsed radio emission \citep{BBL-20}.   

Compared to compressive magnetosonic waves, \A waves do not suffer strong attenuation via non-linear shock formation \citep[e.g.][]{Whitham,CH74} or collisionless dumping \citep[e.g.][]{MT64,Barnes66,Kellogg20}, and hence can traverse large distances from their place of origin. However, they may eventually suffer the parametric instability \citep[e.g.][ for non-relativistic \A waves]{GO62,SG69,Gold78,Derby78} and \citep[][ for relativistic \A waves]{II24},  which provides a mechanism of in-situ plasma heating via the generation of Alfv\'enic turbulence.  

The basic theory of planar linear and non-linear \A waves in ideal relativistic magnetohydrodynamics (RMHD) was developed by \citet{AnilePenn87} and \citet{ssk-aw}. In \citet{ssk-aw}, it was shown that generally  these waves are elliptically polarised, meaning that the tips of the transverse field components trace arcs of ellipses. However, it was also shown that there exist inertial frames, where the elliptical polarisation reduces to circular. This includes the Hoffmann-Teller frame, where both the electric field and the phase speed of the \A wave vanish. \citet{DZ07} and later \citet{II24} presented an RMHD solution describing circularly polarised \A waves, which they treated as the solution describing \A waves in the rest frame of undisturbed plasma\footnote{Apparently, \citet{II24} were unaware of the previous studies of relativistic \A waves. }.       

On the other hand, the non-linear \A waves of Force-Free Degenerate Electrodynamics (FFDE, also known as Magnetodynamics), which can be regarded as RMHD in the singular limit of negligible inertia of plasma, have linear polarisation \citep{ssk-ffde}.  In this paper, I present the resolution to this apparent contradiction. Since the rest frame of undisturbed plasma stands out as the uniquely defined inertial frame, particular attention is paid to the properties of RMHD \A waves as observed in this frame. Their properties in other inertial frames, which may also be considered relevant in applications, can be obtained via the Lorentz transformation.            

Section \ref{sec:prelim} outlines the methodology of this study.  Small-amplitude (linear) \A waves of RMHD are described in section \ref{sec:linAW}. The covariant description of large-amplitude \A waves is summarised in section \ref{sec:NLAW}. Large-amplitude (non-linear) \A waves in the Hoffmann-Teller (HT) frame are described in section \ref{sec:AW-HT}.  In section \ref{sec:AW-II}, the frame implicitly introduced by \citet[][DZBL frame]{DZ07} is identified, and some interesting  peculiarities of the \A wave solution in this frame, particularly in the limit of high plasma magnetisation, are discussed. In section   \ref{sec:AW-UPF}, the properties of \A waves in the rest frame of unperturbed plasma are derived. Section \ref{sec:FFDE} describes the fast and \A waves of FFDE in the rest frame of undisturbed plasma.  Section \ref{sec:discussion} reveals the consistency of the \A wave properties in RMHD and FFDE,  explains the limitations of the HT and DZBL  frames, and describes the breakdown of RMHD (and hence FFDE) approximation for high-frequency \A waves due to charge starvation.            

Throughout the paper, the equations are written in Lorentz-Heaviside units.

\section{Preliminaries}
\label{sec:prelim}

When the spatial scale of variations introduced by waves is small compared to the characteristic spatial scale of the unperturbed state, the plane wave model is well suited to describe the local wave properties.  For quasi-linear systems, these waves are solutions to the 1D equations of the form 
\beq
\pder{\vv{V}}{t}+\mathbb{A(\vv{V})}\pder{\vv{V}}{x}=\vv{0}\,.
\label{eq:qlin-sys}
\eeq 
Finding small-amplitude (linear) plane waves allowed by hyperbolic systems of quasi-linear differential equations, and determining their properties, reduces to solving the eigenvalue problem 
\beq
(\mathbb{A}(\vv{V}_0)-\mu\mathbb{E})\vv{R}=\vv{0}\,
\eeq
for the characteristic matrix of the system linearised about the uniform state \citep[e.g.][]{JT64,Whitham}.  Its eigenvalues $\mu_{(j)}$ determine the phase speeds of linear waves, whereas the corresponding eigenvectors $\vv{R}_{(j)}$ reveal the relation between the variations of the state parameters (the dependent variables $V_i$ of the differential equations ) driven by these waves.  Namely, for the $j$th hyperbolic wave, 
\beq
\vv{\delta V}_{(j)}(\vv{x},t)\, = f(\phi(x,t)) \vv{R}_{(j)}\,,
\eeq
 where $\phi(x,t)=-\mu_{(j)} t+x$ is the phase function describing the phase propagation, and $f(\phi)$ defines the wave profile. The linear waves do not have to be the Fourier harmonics 
 \beq
 \delta\vv{V}_{(j)}=A\vv{R}_{(j)} e^{ik\phi(x,t)}\,, 
 \eeq
 where $k$ is the wavenumber,  but if they are, then the eigenvectors determine the relation between the amplitudes of the harmonics describing the perturbations of individual state variables.  

These results for linear waves can also be used to construct non-linear wave solutions whose local properties are the same as those of the corresponding linear waves.  These so-called simple waves are defined by the condition 
\beq
\vv{V}(x,t)=\vv{F}(\phi(x,t))\,.
\eeq
Substituting this vector function into \eqref{eq:qlin-sys} shows that $\phi$, and hence $\vv{V}$ must be constant along the characteristics of $j$th hyperbolic wave defined by the equation
\beq
\oder{x}{t} = \mu_{(j)}(\vv{V}(\phi)) \,.
\label{eq:char}
\eeq
Thus, the number of simple wave solutions is the same as the number of linear hyperbolic waves of the system. Moreover, equation \eqref{eq:char} implies that their characteristics are straight lines. Thus, $\phi(x,t)$ still has the meaning of the phase function, and the phase speed of the wave is still given by the eigenvalue of the characteristic matrix of system. Hence, the motion of each phase plane is still given by $x =x_0+\mu_{(j)}t$, but the value of $\mu_{(j)}$ may vary from plane to plane, leading to the phenomenon of non-linear steepening or flattening of simple waves.  Furthermore, $\vv{V}_{(j)}(\phi)$ must satisfy the condition 
\beq
\oder{\vv{V}_{(j)}}{\phi} = f(\phi) \vv{R}_{(j)}(\vv{V}(\phi))\,.
\label{eq:gRI}
\eeq  
In this equation, $R_{(j)}$ is one particular eigenvector from the eigenspace corresponding to $\mu_{(j)}$, and $f(\phi)$ is the function defining the wave profile.   Thus, like in the case of linear waves, the variations of the state parameters with the wave phase $d\vv{V}$ are still proportional to the eigenvector.    The integrals of \eqref{eq:gRI} are called generalised Riemann invariants \citep{JT64}.

When the eigenvalue is a single solution of the eigenvalue problem, the corresponding eigenvectors form a one-dimensional vector space. Since only the direction of the eigenvector matters for the variations of the state parameters in the wave, any one of them is defining. When an eigenvalue is a multiple solution of the eigenvalue problem, the corresponding space eigenvectors is multidimensional. Such degenerate waves exhibit greater variety of their properties compared to non-degenerate ones because any linear combination of the basis vectors of their eigenspace describes an allowed wave.  A well-known example is vacuum electrodynamics, where the speed of light is an eigenvalue of double multiplicity. As a result, monochromatic electromagnetic waves can be linearly, circularly, and elliptically polarised.                                      

The basic theory of planar \A waves in ideal relativistic magnetohydrodynamics (RMHD) was developed in \citep{AnilePenn87,ssk-aw}.  They used covariant formulation of RMHD, where the electromagnetic field is described by the 4-vector  $b_\alpha=\frac{1}{2}\eta_{\alpha\beta\gamma\delta}u^{\beta}F^{\gamma\delta}$, where $\eta_{\alpha\beta\gamma\delta}$, $u^\alpha$, and $F^{\alpha\beta}$ are, respectively, the Levi-Civita alternating tensor, the 4-velocity of plasma, and the Maxwell tensor of electromagnetic field \citep{Fr74,Anile89}.  In the 3+1 formulation, the properties of relativistic \A waves can be described in terms of the usual electric, magnetic, and velocity vector fields. These are related to the time and space components of $b^\alpha$ measured in this frame by 
\beq
\vv{B}=u^0\vv{b}-b^0\vv{u}\,,
\label{eq:BEub}
\eeq
\beq
\vv{E} = -\vpr{u}{b}\,,
\label{eq:E}
\eeq 
\beq
b^0=\spr{B}{u}\,,
\eeq
\beq
b^2=b_\nu b^\nu = B^2-E^2\,,
\eeq
where $u^0$ and $b^0$ are the time components, and $\vv{u}$ and $\vv{b}$ are the space components of the 4-velocity vector $u^\nu$ and $b^\nu$ respectively \citep{ssk-godun99}.

In the theory of relativity, wave's phase function must be a scalar field of spacetime. For plane waves, this implies that the spacetime vector $x^\nu$ must enter it in the form $\phi_\nu x^\nu$, where $\phi_\nu$ is a 4-vector of spacetime, containing the information on the orientation of the wave fronts in space and the phase speed.  The basic form is $\phi(x^\nu)=\phi_\nu x^\nu$, which implies that $\phi_\nu = \nabla_\nu\phi$.  
In the 3+1 form with $x^\nu = (t,\vv{x})$ and $\phi^\nu=(\mu,\vv{n})$, where as before $\vv{n}$ is a unit vector, it can be written as   
\beq
\phi(\vv{x},t)=-\mu t +\spr{n}{x}\,.
\label{eq:phase-fun}
\eeq
It is easy to verify that Lorentz transformation does not preserve the spatial part of $\phi^\nu$ as a unit vector. However, without any loss of information, one can simply rescale $\phi(\vv{x},t)$ to make it a unit vector.    

For a linear harmonic wave of the frequency $\omega$ and the wave vector $\vv{k}$, as measured in the selected inertial frame $\Sigma$, the phase function 
\beq
\phi_h(\vv{x},t)=-\omega t +\spr{k}{x})
\label{eq:phase-funm}
\eeq
is a better option as it requires no rescaling.

\section{Linear  Alfv\'en waves} 
\label{sec:linAW}

Linear waves are solutions to linearised equations governing the dynamics of infinitesimally small perturbations. The phase speeds of linear \A waves in Relativistic Magnetohydrodynamics are solutions to the equation
\beq
{\cal E} a^2 -{\cal B}^2=0\,,
\label{eq:cov-speed}
\eeq 
where $a=u^\nu\phi_\nu$, ${\cal B}=b^\nu\phi_\nu$, and  ${\cal E}=w+b^2$, where $w$ is the relativistic enthalpy of plasma \citep{AnilePenn87,Anile89}.  The corresponding eigenvectors show that a) the pressure $p$ and the rest-mass density measured in the comoving frame $\rho$ of plasma are not perturbed by \A waves, b)  $b^2 =b_\nu b^\nu$, $a$, ${\cal B}$ and hence the phase speed $\mu$ are also unchanged, 
\beq
\delta b^2 = \delta a = \delta\cB = \delta\mu=0\,,
\label{eq:invariants}
\eeq
and c) the perturbations of plasma velocity and electromagnetic field are related via the equation     
\beq
\delta u^\alpha = \frac{a}{\cal B} \delta b^\alpha
\label{eq:cov-var}
\eeq    
\citep{ssk-aw}.

In the rest frame of the background plasma, $u^\alpha=(1,\vv{0})$, $b^\nu=(0,\vv{B}_0)$, and hence $a=-\mu$ and $\cB=\vv{n}\cdot\vv{B}_0=B_{0,n}$. These allow to write equations \eqref{eq:cov-speed}  and \eqref{eq:cov-var} as 
\beq
\mu=\pm\frac{B_{0,n}}{\sqrt{\cE}}\,,
\label{eq:lin-mu}
\eeq  
and
\beq
\vv{\delta u} = \mp\frac{1}{\sqrt{\cE}} \vv{\delta B}\,.
\eeq
Since in this frame $\delta b^2=\spr{B}{\delta B}$,  the property  $\delta b^2=0$ reads 
\beq
\vv{\delta B}\perp \vv{B}_0\,.
\eeq
Finally, equation \eqref{eq:E} yields 
\beq
\vv{\delta E} = -\vpr{\delta u}{B}_0\,.
\eeq
Because $\delta B \ll B$, the Lorentz factor of the flow is almost unity, and hence in these equations $\vv{\delta u}$ can be replaced with $\vv{\delta v}$, where $\vv{v}$ is the 3-velocity of plasma\footnote{Throughout this paper we adopt the common in physics notation for the magnitude of a vector $B\equiv||\vv{B}||$.}. 

If we split the vectors in components aligned with $\vv{n}$ and perpendicular to it, e.g. $\vv{B}_0=\vv{B}_{0,n}+\vv{B}_{0,\perp}$, then 
\beq
\vv{\delta u}_n=\vv{\delta B}_n =\vv{\delta E}_n=0\,,
\eeq
and
\begin{align}
\label{eq:dudB}
\vv{\delta u}_\perp &= \mp\frac{1}{\sqrt{\cE}} \vv{\delta B}_\perp\,,\\
\label{eq:dEdB}
\vv{\delta E}_\perp &= -\mu \vpr{n}{\delta B}_\perp\,.
\end{align}
Thus, the linear relativistic \A waves in the plasma frame are described by the same equations as in the Newtonian MHD apart from the mass density $\rho$ being replaced by $\cE$.  In particular, they can be described as purely transverse waves.   

The plasma magnetisation can be quantified using the spacetime scalar $\sigma=b^2/w$. In the rest frame of the background plasma, $b^2=B^2$ and the phase speed \eqref{eq:lin-mu} reads 
\beq
\mu = \pm\frac{B_n}{B}\frac{1}{\sqrt{1+\sigma^{-1}}} \,.
\eeq  

The allowed polarisation of linear \A wave depends on the value of $B_{0,\perp}$. 

(i) If $B_{0,\perp}\ne 0$, then $\vv{\delta B}$ is normal to both $\vv{B}_{0,n}$ and $\vv{B}_{0,\perp}$, and hence the wave is linearly polarised. 

(ii) $B_{0,\perp}= 0$ is a degenerate case, where the phase speed of \A wave coincides with the phase speed of one of the magnetosonic waves, the fast wave if the \A speed $c=B_0/\sqrt{\cE}$ is higher than the sound speed, and the slow wave otherwise \citep{ssk-godun99}.  The corresponding space of eigenvectors is two-dimensional with the basis vectors describing linearly polarised transverse waves satisfying the orthogonality condition  $\vv{\delta B}\cdot\vv{B}_0=0$. If these waves are monochromatic, $\vv{\delta B}= \vv{\delta B}_m e^{i\phi_m}$ where $\phi_m=-\omega t+\spr{k}{r}$, then depending on the ratio of their amplitudes and phase difference their linear combination describes linearly,  circularly,  or elliptically polarised wave.  However, one is free to choose any vector function $\delta \vv{B}(\phi(\vv{r},t))$ which satisfies the orthogonality condition. For example, $\vv{\delta B}$ does not have to rotate one way about $\vv{B}$, but may switch the direction of rotation anywhere before completing a full turn, resulting in arc-polarised wave \citep[cf.][]{MT24}. 

\section{Non-linear Alfv\'en waves} 
\label{sec:NLAW}

To apply the general theory of simple waves, one only needs to replace in equations (\ref{eq:invariants},\ref{eq:cov-var}) perturbations $\delta V$ with differentials $dV=\dot{V}d\phi$, where $\dot{V}=dV/d\phi$, and obtain the differential equations 
\beq
 d(b^2) = da = d\cB =d\mu=0\,,
\label{eq:invariants1}
\eeq
and
\beq
du^\alpha = \frac{a}{\cB} db^\alpha\,.
\label{eq:cov-var1}
\eeq  

Thus, $b^2$, $a$, $\cB$, $\mu$, and $u^\alpha-(a/\cB) b^\alpha$ are generalised Riemann invariants of simple \A waves. The fact that the phase speed $\mu$ is constant means that large-amplitude \A waves do not experience non-linear steepening. Instead their phase profiles remain unchanged during propagation.      

To describe the solution in terms of the spatial vectors $\vv{E}$, $\vv{B}$ and $\vv{u}$, one obviously needs to introduce the inertial frame where these spatial vector fields are measured.

\section{Non-linear Alfv\'en wave in the Hoffmann-Teller frame} 
\label{sec:AW-HT}

The fact that the phase speed of \A waves does not depend on the wave phase (invariant) allows a frame where the electric field vanishes everywhere. Originally, it was proposed by Hoffmann and Teller for analysis of oblique magnetohydrodynamic shocks \citep{HT50}.  Here, the existence of such a frame for \A waves is shown using a similar approach. 

The first step is to select a locally comoving frame $\Sigma'$ (called here the base plane), in which the electric field vanishes at a particular phase plane  $\phi(x^\nu)=\phi_0$.  In this frame, the phase speed is given by equation \eqref{eq:lin-mu} applied at the base plane,
\beq
\mu' = \pm\frac{B'_{n'}(\phi_0)}{\sqrt{{\cE}(\phi_0)}}\,.
\eeq

Next, we introduce another frame $\Sigma_{HT}$, which moves relative to $\Sigma'$ with the velocity $\vv{\beta} =\alpha\vv{B}'(\phi_0)$, where $\alpha$ is such that wave's phase speed in this frame vanishes. In this new frame, the magnetic and electric fields of the base plane are the same as in $\Sigma'$, $\vv{B}(\phi_0)=\vv{B}'(\phi_0)$,  $\vv{E}(\phi_0)=0$. The components of the phase vector defined in $\Sigma'$ as $\phi^\nu = (\mu',\vv{n}')$ now read 
\beq
    \phi^\nu=S \left( \frac{\gamma}{S}(\mu'-\spr{\beta}{n'}), \vv{n}\right)\,,
\eeq
where $\vv{n}=\vv{S}/S$ and  $\vv{S}=\vv{n}' +\vv{\beta}[-\gamma\mu'+(\gamma-1)\spr{\beta}{n'}/\beta^2]$. Hence, the phase speed vanishes if $\mu'-\spr{\beta}{n'}=0$, leading to $\alpha = \pm1/\sqrt{{\cE}(\phi_0)}$ and  
\beq
\vv{\beta}=\pm\vv{B}(\phi_0)/\sqrt{{\cE}(\phi_0)}\,.
\label{eq:HT-beta}
\eeq 

In the frame $\Sigma_{HT}$, all the fields are time-independent, and in the base plane the electric field vanishes. This can be used to show that $\vv{E}=0$ throughout the whole wave. Firstly, the vanishing time derivative of $\vv{B}$ ensures $\vcurl{E}=0$ and hence any contour integral of $\vv{E}$ vanishes as well. If one considers a rectangle with two sides parallel to $\vv{n}$, and the other two sides perpendicular to it, then the total contribution from the sides parallel to $\vv{n}$ vanishes simply because the solution depends only on the coordinate measured along $\vv{n}$. If one of the perpendicular sides belongs to the base plane, then its contribution vanishes too, and so must be the contribution of the other side, yielding $\vv{E}_\perp=0$ . Secondly, because $\spr{E}{B}=0$ (see equations \ref{eq:BEub} and \ref{eq:E}) and $\vv{B}_n\ne \vv{0}$, one may conclude that $\vv{E}_n=0$ as well. Thus in the frame $\Sigma_{HT}$,  $\vv{E}=0$ throughout.  It is for this reason, this frame is referred to as the Hoffmann-Teller (HT) frame of \A wave.

In the HT frame,  $b^\alpha=(0,\vv{B})$, $b^2=B^2$, $\phi_\nu=(0,\vv{n})$, $a=\spr{u}{n}=u_n$, and ${\cal B}=\spr{B}{n}=B_n$. Since  $b^2$ is a wave invariant, one immediately concludes that $B^2$, and hence of $B_\perp$, do not depend on the wave phase,
\beq
B^2, B_\perp =\mbox{const} \,.
\eeq 
It is also easy to see that 
\beq
v^2, v_\perp =\mbox{const} \,.
\eeq
Firstly, the invariance of $a$ now implies that $u_n$ is constant. Secondly, because $b^0=0$, the time component of equation \eqref{eq:cov-var1} reads $du^0=0$, and hence both $u^0$ and $v^2$ are constant.   

The spatial component of \eqref{eq:cov-var1} now reads 
\beq
d\vv{v} - \frac{v_n}{B_n}\vv{B} =0\,,
\eeq
and hence
\beq
\vv{v} - \frac{v_n}{B_n} \vv{B} = \vv{C}\,,  
\eeq 
where $\vv{C}$ is a vector constant.  Because in the base plane 
$$
\vv{v}=-\vv{\beta} = \mp\vv{B}/\sqrt{\cE}\,,
$$ 
one immediately finds that $\vv{C}=0$, and 
\beq
\vv{v}=\mp\frac{1}{\sqrt{\cE}}\vv{B}
\label{eq:HT-vB}
\eeq
in every phase plane of the wave. Thus, in the HT plane the properties of non-linear relativistic \A waves are remarkably similar to the properties of their Newtonian counterparts.  

The invariance of $B_\perp$ implies that the tip of $\vv{B}_\perp$ traces a circular arc. None of the above equations      
puts any constraints on how exactly it does this, apart from the condition the $\vv{B}_\perp(\phi)$ is a differentiable function\footnote{In fact, it is easy to show that week solutions are also allowed.}.  For example, the tip of $\vv{B}_\perp$ may trace arcs of the angular size less than $2\pi$, which corresponds to non-linear arc-polarised \A waves. 

In addition to the HT frame, there are infinitely many other frames where the phase speed of the \A wave vanishes too. In these frames, $\vv{E}$ does not vanish, $B_\perp$ (and $v_\perp$) is no longer constant,  and the tip of  $\vv{B}_\perp$ (and $\vv{E}_\perp$ ) traces elliptical arcs instead of circular ones \citep{ssk-aw}.    

\section{Non-linear Alfv\'en waves in the DZBL frame} 
\label{sec:AW-II}

In the frame moving relative to the HT frame with the velocity 
\beq
\vv{\beta}_n=\mp\vv{B}_n/\sqrt{\cE}\,,
\eeq 
the phase speeds of the \A wave
\beq
\mu=\pm\frac{B_n}{\sqrt{\cal E}}\,. 
\label{eq:mu-ii}
\eeq
This is the same as for linear waves in the plasma frame (see equation \ref{eq:lin-mu}). In addition, the  Lorentz transformation also yields
\beq
\vv{v}_n=0, \, \vv{E}_n=0\,,
\label{eq:vEn}
\eeq
\beq
\vv{v}_\perp=\mp \frac{\vv{B}_\perp}{\sqrt{\cal E}} \,,
\label{eq:vBp}
\eeq
and
\beq
\vv{E}_\perp=-\vv{v}_\perp\!\times\!\vv{B}_n=-\mu \vpr{n}{B}_\perp\,.
\label{eq:EBcp}
\eeq
Moreover, the magnitudes of $\vv{B}_\perp$, $\vv{E}_\perp$ and $\vv{v}_\perp$ still remain phase-independent, and hence their tips still trace circular arcs.  Thus, the properties of non-linear \A waves in this frame appear to be very similar to the properties of their Newtonian counterparts.   This  frame (called here the DZBL  frame) was implicitly adopted by \citet{DZ07}, and later by \citet{II24} in their analysis of the parametric instability of circularly polarised monochromatic  \A waves.

\begin{figure*}
\centering
 \includegraphics[width=0.7\columnwidth]{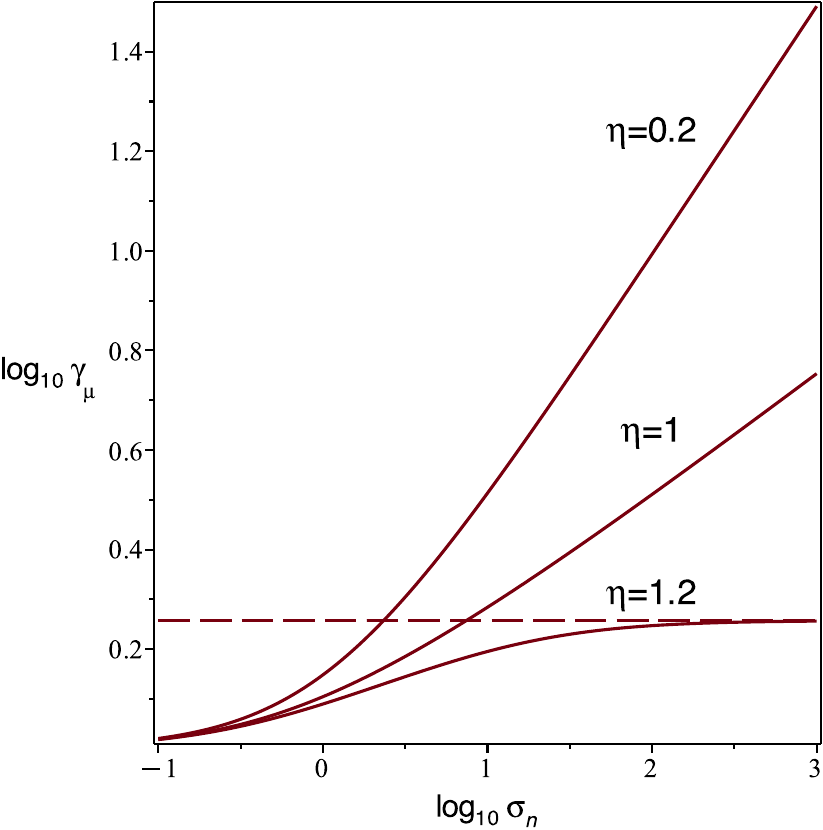}
 \includegraphics[width=0.7\columnwidth]{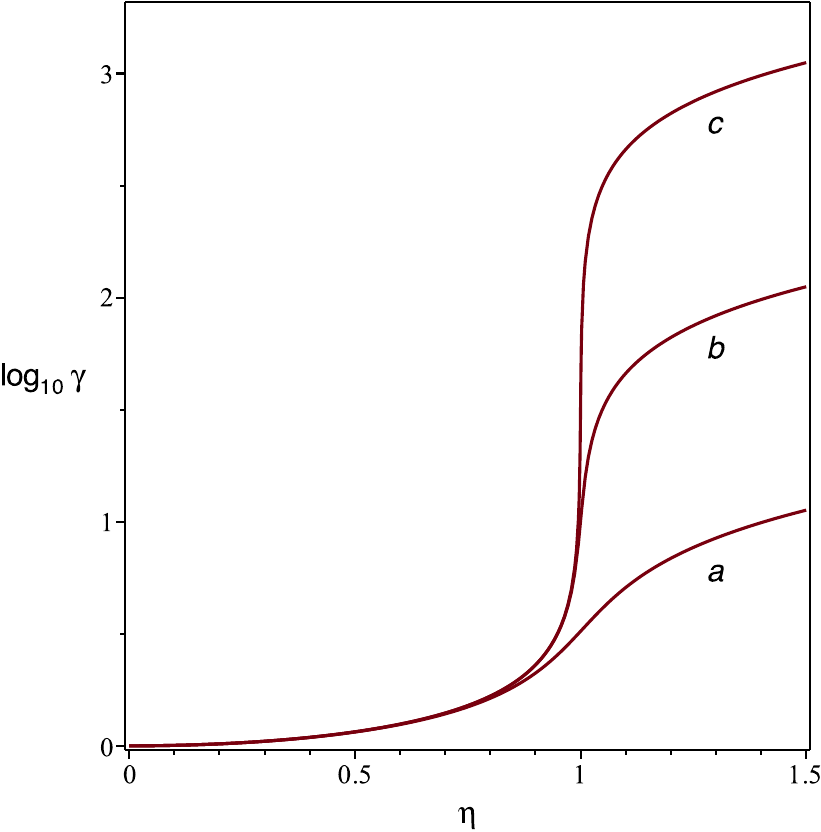}
 \caption{\A waves in Ishizaki-Ioka frame. Left panel: Lorentz factor of the phase speed as a function of the magnetisation parameter $\sigma^{}_n$.
 Right panel: Lorentz factor of plasma motion as a function of the wave amplitude $\eta$ for $\sigma^{}_n=10^2$ (a), $10^4$ (b),  and $10^6$ (c). }
\label{fig:gamA}
\end{figure*}

Although equation \eqref{eq:mu-ii} does look very similar to its Newtonian counterpart, its appearance is deceptive.  This is because 
\beq
{\cal E}=w+B^2-E^2\,,
\eeq 
where 
\beq
E^2=\mu^2 B_\perp^2
\label{eq:Ec0B}
\eeq 
also depends on $\mu$. Straightforward calculations show that equation \eqref{eq:mu-ii} reduces to the quadratic equation\footnote{Equation \eqref{eq:Y} is equivalent to equations (85) in \citet{DZ07} and (16) in \citet{II24}.}  
\beq
\gamma_\mu^4-(1+\sigma^{}_n(1-\eta^2))\gamma_\mu^2-\sigma^{}_n\eta^2=0\,,
\label{eq:Y}
\eeq
where $\gamma_\mu=(1-\mu^2)^{-1/2}$,  $\eta=B_\perp/B_n$ and 
\beq
\sigma^{}_n = \frac{B_n^2}{w}=\frac{\sigma}{1+\eta^2}
\eeq
is the magnetisation parameter introduced by \citet{II24}. It is easy to show that as $\sigma^{}_n \to \infty$
\beq
 \gamma_\mu^2 \simeq 
 \begin{cases}
 (1-\eta^2) \sigma^{}_n , & \etext{if } \eta <1 \\
 \sigma_n^{1/2}, & \etext{if } \eta =1 \\
 1/(1-\eta^{-2}), & \etext{if } \eta >1 
  \end{cases}
  \label{eq:assY}
\eeq  
Thus, there are two qualitatively different regimes -- one for $\eta<1$ and another for $\eta>1$. This division is illustrated in the left panel of figure \ref{fig:gamA}. 
 
Equation \eqref{eq:vBp} helps to understand why $\eta=1$ is a critical value. It shows that the speed of plasma motion in the \A wave 
\beq
  v=\eta\mu \,.
  \label{eq:v-mu}
\eeq 
Hence if $\eta<1$, then $v<1$ (less than the speed of light) for any $\mu\le 1$, thus allowing highly relativistic wave speeds. If however $\eta>1$, then the condition $v<1$ requires $\mu<\eta^{-1}$.   

Using equation \eqref{eq:v-mu} to replace $\mu$ with $v$ in equation \eqref{eq:mu-ii}, one can derive the equation for the Lorentz factor of plasma motion in the DZBL solution,   
\beq
\gamma^4+(\sigma^{}_n(1-\eta^2)-1)\gamma^2-\sigma^{}_n=0\,.
\label{eq:X}
\eeq 
It is easy to see that as $\sigma^{}_n \to \infty$
\beq
 \gamma^2 \simeq 
 \begin{cases}
 (1-\eta^2)^{-1}, & \etext{if } \eta <1 \\
 \sigma_n^{1/2}, & \etext{if } \eta =1 \\
 \sigma^{}_n (\eta^2-1), & \etext{if } \eta >1 .
  \end{cases}
  \label{eq:assX}
\eeq  
The right panel of figure \ref{fig:gamA} illustrates how rapidly the plasma motion becomes ultra-relativistic near $\eta=1$ when the magnetisation $\sigma_n\gg 1$.   

The high value of the plasma Lorentz factor in large-$\eta$ waves invites comparing wave's kinetic energy  $\varepsilon_k \approx w \gamma(\gamma-1)$  to its electromagnetic energy $\varepsilon_{m}=(B_\perp^2+E_\perp^2)/2$. Using equations \eqref{eq:Ec0B},\eqref{eq:assY}, \eqref{eq:assX}, one finds that
\beq
\frac{\varepsilon_{m}} {\varepsilon_k}\simeq \frac{1}{2}\frac{\eta^2+1}{\eta^2-1}  \etext{as} \sigma^{}_n\to\infty\,.
\label{eq:ke-la}
\eeq 
Thus even in highly magnetised plasma the energies can be comparable, with their ratio actually independent on $\sigma^{}_n$. 

In contrast, in the case of small-$\eta$ waves $\varepsilon_k \approx {w\eta^2}/{2}$, and
\beq
  \frac{\varepsilon_{m}} {\varepsilon_k}\simeq  {2}\sigma^{}_n  \etext{as} \sigma^{}_n\to\infty\,.
 \label{eq:ke-sa}
\eeq
Thus in highly-magnetised plasma kinetic energy of such waves is very small compared to their electromagnetic energy. 

Since $\vv{v}_n=0$,  one may think that the DZBL  frame is the rest frame of background unperturbed state. This interpretation is even more convincing in the case of circularly polarised \A waves, because in this case the averaged over the wave period value of the transverse velocity also vanishes, $<\!\!\vv{v}_\perp\!\!\!>=0$.  However, there are only two ways an \A wave can be connected to the uniform state of unperturbed plasma at its wave front, either continuously or via an \A discontinuity. In both these cases the speed of the transverse motion of plasma must be the same on the both sides of the front  \citep{ssk-aw}.  Hence in the DZBL  frame, the plasma speed in the unperturbed state does not vanish, and so this frame is actually not the rest frame of unperturbed plasma.

\section{Non-linear Alfv\'en in the rest frame of unperturbed plasma}
\label{sec:AW-UPF}

 In the rest frame of unperturbed plasma $u^\nu_0=(1,\vv{0})$ and $b^\nu_0=(0,\vv{B}_0)$, where $\vv{B}_0$ is the unperturbed magnetic field, and the 4-gradient of adjusted phase function is $\phi^\nu=(\mu,\vv{n})$.  This sets the values of the wave invariants  
\begin{align}
\label{eq:b2}
b^2&=B_0^2,\\
a&=-\mu,\\
\label{eq:cB}
\cB&=B_{0,n}\,,
\end{align}
whereas equation \eqref{eq:cov-speed} yields the phase speeds
\beq
\mu=\pm\frac{B_{0,n}}{\sqrt{\cE}}\,,
\label{eq:mu-rmhd}
\eeq
where $\cE=w_0+B_0^2$. Equation \eqref{eq:cov-var1} now reads 
\beq
d(u^\nu \pm \frac{1}{\sqrt{\cE}}  b^\nu)=0\,.
\label{eq:cov-var1a}
\eeq
It's general solution is  
\beq
u^\nu \pm \frac{1}{\sqrt{\cE}}  b^\nu =c^\nu\,,
\label{eq:unu-bnu}
\eeq
and the initial condition at the leading front allows us to fix the constant 4-vector $c^\nu=(1,\pm\vv{B}_0/\sqrt{\cE})$. Hence, 
\beq
u^0=1\mp\frac{b^0}{\sqrt{\cE}}\,, 
\label{eq:u0}
\eeq 
and
\beq
\vv{u}=\pm\frac{1}{\sqrt{\cE}} (\vv{B}_0-\vv{b})\,.
\label{eq:uBb}
\eeq
Formally, these equations allow us to find the functions $u^\nu(\phi)$ given the  functions $b^\nu(\phi)$. However, the functions $b^\nu(\phi)$ themselves must be consistent with equations \eqref{eq:b2} and  \eqref{eq:cB}, as well as the constraint 
\beq
u_\nu b^\nu=0 \,,
\label{eq:ubc}
\eeq
which follows from definition of $b^\nu$. 

Substituting \eqref{eq:u0}  and \eqref{eq:uBb} into \eqref{eq:ubc}  yields
\beq
b^0=\mp\frac{1}{\sqrt{\cE}}\left(B_0^2-\spr{b}{B_0}\right) \,,
\label{eq:b0}
\eeq
 whereas substituting them into \eqref{eq:cB} yields
\beq
b_n=B_{0,n}+\mu b^0 \,,
\label{eq:bn}
\eeq
Finally, substituting \eqref{eq:b0} and \eqref{eq:bn} into \eqref{eq:b2}  allows us to find the equation constraining $\vv{b}_\perp(\phi)$.  After little algebra, all the equations constraining $b^\nu(\phi)$ can be written as   
\beq
    \wb^0=\frac{\mu}{1-\mu^2}\spr{\eta^{}_0}{\wb_\perp}\,,
    \label{eq:b0a}
\eeq
\beq
\wb_n=\frac{\mu^2}{1-\mu^2}\spr{\eta^{}_0}{\wb_\perp}\,,
\eeq
\beq
\wb_\perp^2-\frac{\mu^2}{1-\mu^2}\spr{\eta^{}_0}{\wb_\perp}^2 +
2(1+\frac{\mu^2}{1-\mu^2}) \spr{\eta^{}_0}{\wb_\perp}=0\,,
\label{eq:bpa}
\eeq
where
$$
\wb^\nu=(b^\nu-b^\nu_0)/B_{n,0}\,,
$$
is the perturbation of $b^\nu_0$ normalised to the value of $B_{0,n}$, and 
$$
\vv{\eta}^{}_0=\vv{B}_{0,\perp}/{B_{0,n}}\,.
$$ 

Introducing the Cartesian basis $\{\ort{x},\ort{y},\ort{z}\}$, where $\ort{x}=\vv{n}$, $\ort{y}$ is aligned with $\vv{\eta}_0$ and $\ort{z}=\ort{x}\times\ort{y}$, allows us to reduce equations \eqref{eq:b0a}-\eqref{eq:bpa} to
\beq
\wb^0=\frac{\mu\eta^{}_0}{1-\mu^2} \wb_y\,,
\eeq
\beq
\wb_x=\frac{\mu^2\eta^{}_0}{1-\mu^2}\wb_y\,.
\eeq
\beq
(1-\epsilon^2)\wb_y^2 +\wb_z^2+\frac{2\eta^{}_0}{1-\mu^2}\wb_y=0\,,
\label{eq:ellipse-b}
\eeq
where  
\beq
\epsilon=\frac{|\mu|\eta^{}_0}{\sqrt{1-\mu^2}} \,.
\label{eq:eccentricity}
\eeq
Obviously, equation \eqref{eq:ellipse-b} describes an ellipse of eccentricity $\epsilon$ and major axis aligned with the y axis. 
As the wave phase changes, the tip of $\vv{b}_\perp(\phi)$ traces an arc of this ellipse. How it does this is not determined by the solution and remains free from any constraints.    

\begin{figure}
\centering
 \includegraphics[width=\columnwidth]{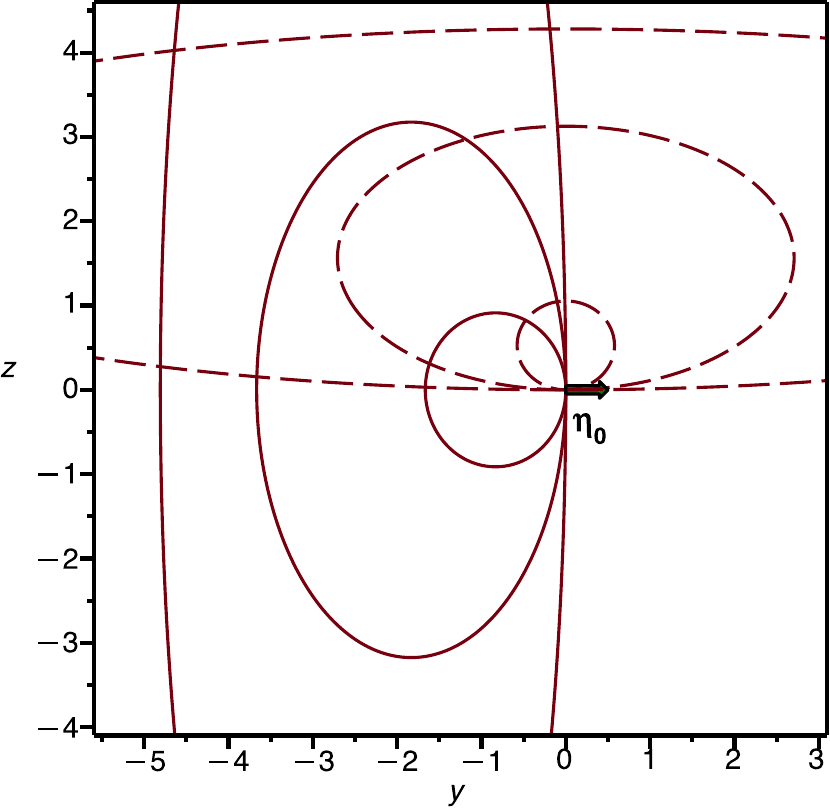}
 \caption{Ellipses traced by the tips of $\vv{\wB}_\perp$ (solid lines), and $\vv{\wE}_\perp$ (dash lines) for \A waves propagating through plasma with $\sigma^{}_0=1,10$ and 100, and $\eta^{}_0=0.5$. As the value of $\sigma^{}_0$ increases, so does the ellipse size. For $\sigma^{}_0=100$, only small sections of the B and E ellipses can be seen, and they are almost straight.  The arrow shows the vector $\vv{\eta}^{}_0=\vv{B}_{0,\perp}/B_{0,n}$. }
\label{fig:ellipses}
\end{figure}

Combining the last results with equations \eqref{eq:u0} and \eqref{eq:uBb}, the components of plasma's 4-velocity can be written as 
\beq
u^0=1-\frac{\mu^2\eta^{}_0}{1-\mu^2} \wb_y\,,
\eeq 
\beq
u_x=-\mu \frac{\mu^2\eta^{}_0}{1-\mu^2} \wb_y\,,
\label{eq:ux}
\eeq
\beq
\vv{u}_\perp=-\mu \vv{\wb}_\perp \,.
\label{eq:up}
\eeq
Equation \eqref{eq:ux} shows that relativistic \A waves are no longer transverse, whereas according to equation \eqref{eq:up}   the tip of $\vv{u}_\perp$ traces an ellipse of the same eccentricity and orientation as the $\wb_\perp$-ellipse.      

Now, one can use equation \eqref{eq:BEub} to show that 
\beq
\wB_x=0\,,
\eeq 
and hence $\vv{B}_n=\vv{B}_{0,n}$, and
\beq
\vv{\wB}_\perp=(1-\epsilon^2)\wb_y\ort{y} + \wb_z\ort{z} \,.
\label{eq:wB}
\eeq
Using the last equation to replace the components of $\vv{\wb}_\perp$ with the components of  $\vv{\wB}_\perp$ in \eqref{eq:ellipse-b} yields 
\beq
\left(\wB_y +\frac{\eta^{}_0}{1-\mu^2} \right)^2 +(1-\epsilon^2)\wB_z^2 =\frac{\eta^{2}_0}{(1-\mu^2)^2}\,.
\label{eq:ellipse-B}
\eeq 
Thus, the tip of $\vv{\wB}_\perp$ traces an ellipse of the same eccentricity as the $\wb_\perp$-ellipse, but its major axis is now aligned with the z axis. 

Similarly, equation \eqref{eq:E} can be used to show that 
\beq
\wE_x = -\mu\eta^{}_0\wB_z \,,
\label{eq:wEx}
\eeq   
\beq
\vv{\wE}_\perp =-\mu\, (\ort{x} \!\times\!\vv{\wB}_\perp)\,,
\label{eq:wEp}
\eeq
and 
\beq
 (1-\epsilon^2)\wE_y^2  + \left(\wE_z - \frac{\eta^{}_0\mu}{1-\mu^2}\right)^2 = \frac{\eta^{2}_0\mu^2}{(1-\mu^2)^2}\,.
\label{eq:ellipse-E}
\eeq 
Thus, the tip of $\vv{\wE}_\perp$ traces an ellipse of the same eccentricity as the $\wB_\perp$-ellipse, but it is smaller by the factor of $\mu$, and its major axis is aligned with the y axis.    

Using equation \ref{eq:wB}, one can relate  plasma's 4-velocity to the wave components of the magnetic field,
\beq
u^0=1-\frac{\mu^2\eta^{}_0}{(1-\mu^2)(1-\epsilon^2)} \wB_y\,,
\label{eq:gam1}
\eeq 
\beq
u_x=-\mu \frac{\mu^2\eta^{}_0}{(1-\mu^2)(1-\epsilon^2)} \wB_y\,,
\label{eq:ux1}
\eeq
\beq
u_y=-\frac{\mu}{1-\epsilon^2} \wB_y \,.
\label{eq:uy}
\eeq
\beq
u_z=-\mu \wB_z \,.
\label{eq:uz}
\eeq

\begin{figure*}
\centering
 \includegraphics[width=0.8\columnwidth]{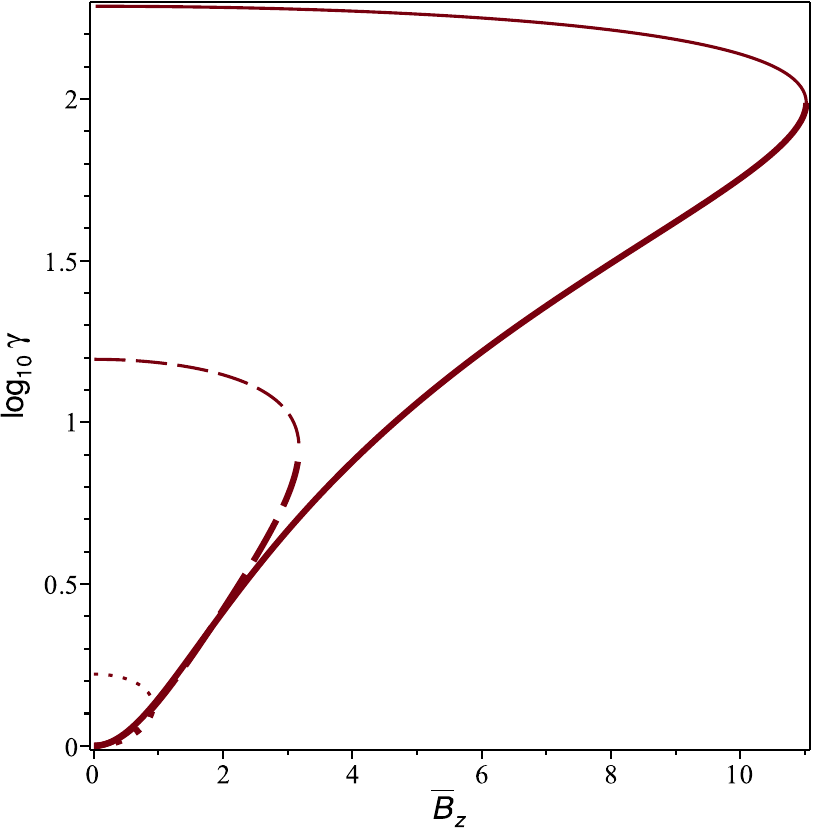}
 \includegraphics[width=0.8\columnwidth]{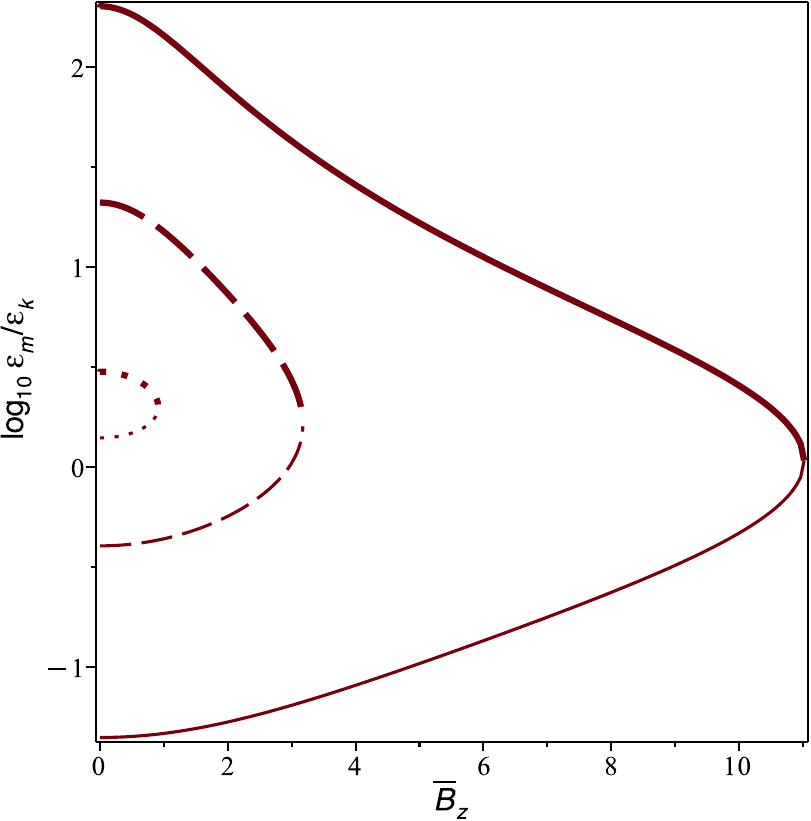}
 \caption{Lorentz factor of plasma motion ( left panel ) and the ratio of wave's electromagnetic to kinetic energy as functions of $\wB_z$ for $\eta^{}_0=0.5$ and $\sigma^{}_0=1$ (dot line),10 (dash line) and 100 (solid line). The thick line corresponds to the inner half, and the thin line to the outer half of the B-ellipse.  }
\label{fig:energy}
\end{figure*}

Figure \ref{fig:ellipses} illustrates the dependence of the $\wB_\perp$- and $\wE_\perp$-ellipses on $\sigma^{}_0$ for $\eta^{}_0=0.5$.

As $\sigma^{}_0\to 0$, both $\epsilon\to 0$ and $\mu\to 0$. Hence, the $\wB_\perp$-ellipse reduces to a circle of radius $\eta^{}_0$, the well-known Newtonian result. 

As $\sigma^{}_0\to\infty$, 
\beq
\mu^2 \simeq \frac{1}{1+\eta^{2}_0} - \frac{1}{1+\eta^{2}_0}\sigma_0^{-1} + O(\sigma_0^{-2})\,,
\eeq
and the eccentricity of the wave ellipses approaches unity
\beq
\epsilon^2\simeq 1-\frac{1+\eta^{2}_0}{\eta_0^2}\sigma_0^{-1} + O(\sigma_0^{-2})\,.
\eeq
At the vertices of the B-ellipse, 
\beq
|\wB_y| \simeq \frac{1+\eta^{2}_0}{\eta^{}_0} - \frac{1+\eta^{2}_0}{\eta^{3}_0}\sigma_0^{-1}+O(\sigma_0^{-2})\,, 
\eeq
\beq
|\wB_z| \simeq (1+\eta^{2}_0)^{1/2} \sigma_0^{1/2} + O(\sigma_0^{-1/2})\,,
\label{eq:vertice-Bz}
\eeq
and thus only $B_z$ increases without limit.  The corresponding Lorentz factor of the plasma motion driven by the wave   
\beq
\gamma \simeq \sigma^{}_0 + O(1)\,,
\eeq 
and the ratio of magnetic to kinetic energies 
\beq
\frac{\varepsilon_m}{\varepsilon_k} \simeq 1 + O(\sigma_0^{-1})\,,
\eeq
are very high in highly magnetised plasma.

Figure \ref{fig:energy} illustrates the variation of the Lorentz factor and the energy ratio in \A waves $\sigma^{}_0=1,10,100$. As one can see, beyond the vertices (in the direction away from the origin) the Lorentz factor is even higher, and the kinetic energy dominates.  In the inner half of the ellipse, it is the other way around.   It is easy to verify that for small-amplitude waves with $\wB_z \ll \eta^{}_0$ 
\beq
\gamma \simeq 1+\frac{\mu^2}{2}\wB_z^2 \etext{and} \frac{\varepsilon_m}{\varepsilon_k} \simeq 2\sigma^{}_0\,.
\eeq
%

\section{non-linear fast and Alfv\'en waves in FFDE}
\label{sec:FFDE}

The general properties of waves in FFDE were derived in \citep{ssk-ffde}.  Here we present the results in a more compact vector form. 

\subsection{Fast waves}

The phase speeds of fast waves are  
\beq
\mu=\pm 1\,.
\eeq  
The simple wave equations of fast waves yield
\beq
d E_n=0 \,,
\eeq
which shows that $E_n$ is a wave invariant and that the fast wave is transverse;
\beq
d(\vv{E}_\perp +\vpr{n}{B}_\perp) =0 \,,
\eeq
which shows that the vector
\beq
\vv{\xi}=\vv{E}_\perp +\vpr{n}{B}_\perp
\eeq
is also a wave invariant; and   
\beq
\spr{\xi}{dB_\perp}=0\,,
\eeq
which shows that the fast wave is linearly polarised, with $\vv{\wB}_\perp$ being perpendicular to $\vv{\xi}$. Apart from the constraint on their polarisation direction, the fast waves of FFDE are the same as linearly polarised electromagnetic waves in vacuum.  
 
In the rest frame of the unperturbed plasma with the magnetic field $\vv{B}_0=B_{0,n}\vv{n} +\vv{B}_{0,\perp}$ and the electric field $\vv{E}_0=\vv{0}$,  
\beq
E_n=0
\label{eq:En-ffde-rfa}
\eeq
\beq
\vv{\xi}=\vpr{n}{B}_{0,\perp} \,,
\eeq
\beq
\vv{\wB}_\perp\, \parallel\, \vv{B}_{0,\perp}\,,
\eeq
and 
\beq
\vv{\wE}_\perp = -\vpr{n}{\wB}_\perp\,.
\label{eq:Ep-ff-rfa}
\eeq

\subsection{\A waves}

Their phase speeds are given by the equation
\beq
\mu=\frac{(\vpr{E}{B})_{n}}{B^2}\pm\frac{B_n}{B^2}(B^2-E^2)^{1/2}\,,
\label{eq:mu-ff}
\eeq
and the simple wave equations yield  
\beq
d(B^2 -E^2)=0\,,
\eeq 
which shows that the magnetic pressure is phase-invariant;
\beq
d\mu=0\,,
\eeq
which shows that $\mu$ is phase-invariant as well;
\beq
d(\vv{E}_\perp+\mu \vpr{n}{B}_\perp)=0\,,
\eeq 
which shows that the vector
\beq
\vv{\zeta} = \vv{E}_\perp + \mu \vpr{n}{B}_\perp
\label{eq:zeta}
\eeq
is also phase-invariant; 
\beq
\spr{\zeta}{dE_\perp} = 0\,,
\eeq
which shows that in terms of their transverse field components \A waves are also linearly polarised; and
\beq
d(\spr{\zeta}{B_\perp} + B_n E_n)=0\,,
\label{eq:En-ff}
\eeq
which shows that $E_n$ also varies with the wave phase, and hence \A  waves are not entirely transverse.

In the rest frame of the unperturbed plasma with the magnetic field $\vv{B}_0=B_{0,n}\vv{n} +\vv{B}_{0,\perp}$ and the electric field $\vv{E}_0=\vv{0}$,   
\beq
\mu=\pm\frac{B_{0,n}}{B_0}\,,
\label{eq:mu-ffde-rf}
\eeq
\beq
\vv{\zeta}=\mu(\vpr{n}{B}_{0,\perp})\,,
\label{eq:zeta-rf}
\eeq
\beq
\vv{\wB}_\perp \perp \vv{B}_{0,\perp} \,, 
\eeq
\beq
\wE_\perp = -\mu (\vpr{n}{\wB}_\perp) \,.
\label{eq:Ep-ff-rf}
\eeq
and
\beq
\wE_n=-\mu \vv{\wB}_\perp\! \cdot (\vpr{n}{\eta}^{}_0)\,,
\label{eq:En-ff-rf}
\eeq
where as before $\vv{\eta}^{}_{0}=\vv{B}_{0,\perp}/B_{0,n}$.

\subsection{Ambiguity of the rest frame of unperturbed plasma} 

Since the plasma velocity is not present in the equations of FFDE, the notion of the plasma rest frame is inevitably ambiguous.  If in some inertial frame the uniform state adjacent to a fast or \A wave has $\vv{B}=\vv{B}_0$ and $\vv{E}=\vv{0}$, then in any other inertial frame moving relative to it with velocity $\vv{v} = \alpha \vv{B_0}$, the electric and magnetic field remain exactly the same as in the original frame.  However, this is not as much of a problem as it may appear at first sight because in all these frames the properties of fast \A waves are exactly the same. They are still given by equations \eqref{eq:En-ffde-rfa}-\eqref{eq:Ep-ff-rfa}  for fast waves and equations \eqref{eq:mu-ffde-rf}-\eqref{eq:En-ff-rf}  for \A waves.  For any particular wave, the Lorentz transformations change the wave's phase speed $\mu$, phase vector $\vv{n}$, and hence $\vv{\eta}_0$, but in the new frame there will be another wave with exactly the same $\mu$, $\vv{n}$, and all other parameters as the original wave in the old frame.     

\section{Discussion}
\label{sec:discussion}

\subsection{Consistency of \A waves solutions in RMHD and FFDE}

FFDE is the singular limit of RMHD corresponding to $\sigma^{}_0=\infty$ \citep{ssk-ffde}.  
It is easy to see that when $\sigma^{}_0=\infty$, the RMHD phase speed of \A waves measured in the rest frame of unperturbed plasma  given by \eqref{eq:mu-rmhd} reduces to the FFDE phase speed given by \eqref{eq:mu-ffde-rf}. Since the eccentricity of the $\wB_\perp$- and $\wE_\perp$-ellipses of the RMHD \A waves $\epsilon\to1$,  and the length of their major axes diverges as $\sigma^{}_0\to\infty$, the polarisation of the transverse field components becomes linear in this limit, with the same polarisation angle as in FFDE.  The transverse components of the electromagnetic field are related by the same equation (compare \eqref{eq:wEp} with \eqref{eq:Ep-ff-rf}).   Finally, comparing \eqref{eq:wEx} with \eqref{eq:En-ff-rf} shows that the longitudinal component of the wave electric field in RMHD in the limit $\sigma\to\infty$ is given by the same equation as in  FFDE.  Thus, the properties of \A waves in FFDE are fully consistent with their properties in RMHD.     

\subsection{Singularity of the HT and DZBL  frames in the FFDE limit}

The linear polarisation of the FFDE \A waves seems to be in conflict with the properties of the RMHD \A waves in the HT and DZBL  frames, which have arc-polarisation for any value of the magnetisation parameter $\sigma$.  This is due to the extreme nature of these frames in the limit $\sigma\to\infty$. Equation \eqref{eq:HT-vB} tells us that the speed of plasma motion in the HT frame $v\to 1$, the speed of light, as $\sigma\to\infty$. This also applies to the speed of plasma motion in the unperturbed state through which this wave propagates. Thus, the HT frame moves with increasingly relativistic speed relative to the unperturbed plasma.   Ultimately, its speed becomes luminal  when $\sigma=\infty$, and the world line of the HT frame becomes null. The same applies to the DZBL frame when the parameter $\eta\ge1$ because, according to equation \eqref{eq:assX},  the plasma speed in this frame also tends to the speed of light, as $\sigma\to\infty$.  

When $\eta<1$,  the plasma speed in the DZBL frame remains sub-luminal, as $\sigma\to\infty$, and hence the above argument does not apply. However, in this case, the phase speed of \A waves in the DZBL frame turns luminal when $\sigma=\infty$.  In FFDE, this is the degenerate case where $\mu=\pm1$ is a double solution of the dispersion equation. The corresponding vector space of eigenvectors is two-dimensional with the basis vectors describing linearly polarised electromagnetic waves.  Because the superposition of two orthogonal linearly polarised monochromatic electromagnetic waves with the same wavevector gives a circularly polarised wave when they have equal amplitude and $90^o$ phase difference, this suggests that for $\eta<1$, the circularly polarised \A wave described by \citet{II24} in the corresponding DZBL  frame reduces to a circularly polarised electromagnetic wave as $\sigma^{}_0\to\infty$.   

To further strengthen this conclusion, one may use the second part of equation \eqref{eq:EBcp}, which in the limit $\sigma_n= \infty$ reads 
\beq
\vv{E} = \mp \vpr{n}{B} \,,
\eeq  
taking into account that in the DZBL frame $E_n=0$. This is exactly the condition of FFDE degeneracies corresponding to double eigenvalues $\mu=\pm1$ respectively \citep{ssk-ffde}.  One may also use the Lorentz transformation to the frame moving with the same speed as plasma in the \A wave solution by \citet{II24} at the leading front of the wave, which will be the rest frame of undisturbed plasma, and verify that in this frame the wave vector $\vv{n}$ becomes align with the magnetic field of  the undisturbed state when $\sigma_n=\infty$.  This is exactly how the degeneracy condition reads in the rest frame of the undisturbed plasma.

\subsection{Charge-current starvation condition}
\label{sec:charge-starv}

 The value of the magnetisation parameter $\sigma_0$ contains information on the number density of electric charges available in the background plasma, which can be checked against the minimum number density of charges required to support the variable electromagnetic field of \A wave. 
 
 Suppose that the wave is monochromatic in the sense that  $B_z=B_{k}e^{i(-\omega t +kx)}$, where $B_k$ is smaller than the value of $B_z$ at the vertice of the $B$-ellipse. Then, according to equations \eqref{eq:wEx} and \eqref{eq:En-ff}, in the rest frame of the undisturbed plasma,
 \beq
 \vdiv{E} =\oder{E_x}{x} = \omega \eta^{}_0 B_{z} \,,
 \eeq   
and the minimum number density of elementary charges required to sustain  this solution is
\beq
n^{}_{k} = \frac{\omega \eta^{}_0 B_{k} }{e}\,,
\eeq 
where $e$ is the electric charge of electrons. If the undisturbed plasma is a cold electron-positron plasma with magnetisation $\sigma_0$, then the number density of its elementary charges is
\beq
n^{}_0 = \frac{B_0^2}{\sigma_0 m_e}\,.
\eeq
Inside the \A wave, the number density of charges is $n=n_0 \gamma$, where $\gamma$ is the Lorentz factor of the plasma motion in the wave given by equation \eqref{eq:gam1}. This density must exceed $n_{k}$, leading to the condition
\beq
\frac{\omega}{\omega_b}< \frac{\gamma}{\sigma_0 \eta_0 \kappa}\,,
\label{eq:cond1}
\eeq 
where $\omega_b=eB_0/m_e$ is the cyclotron frequency, and $\kappa=B_k/B_0$. If condition \eqref{eq:cond1} is not satisfied, the RMHD approximation breaks down. Since  the plasma frequency $\omega_p=\omega_b/\sqrt{\sigma_0}$, this condition can also be written in terms of the plasma frequency as 
\beq
\frac{\omega}{\omega_p}< \frac{\gamma}{\sqrt{\sigma_0} \eta_0 \kappa}\,. 
\label{eq:cond2}
\eeq   
Except for the factor $\gamma$, this is the same as the current starvation condition derived by \citep{Beloborodov-24}. 

In plasma with $\sigma_0\gg 1$, the B-ellipse is highly elongated and its vertice the value of wave's magnetic field can be extremely high compared to $B_{0,n}$ (see equation \ref{eq:vertice-Bz} ) and $\gamma\simeq\sigma_0$.  For more reasonable wave amplitudes, one may consider only the section of the ellipse where $|\wB_y| \ll \eta_0/(1-\mu^2)$. Then equations \eqref{eq:gam1} and \eqref{eq:ellipse-B} yield the estimate
\beq
\gamma \approx 1+\frac{\kappa^2}{2}\,.
\eeq
Thus this factor is significantly above unity only for large-amplitude waves with $B_k>B_0$, whereas  \citep{Beloborodov-24} analysed only small-amplitude (linear) \A waves.  

A similar check can be made for the wave current
\beq
\vv{j} = -\pder{\vv{E}}{t} +\vcurl{B}\,. 
\eeq
In plasma with $\sigma\gg 1$, one may ignore $B_y$ compared to $B_z$, leading to 
\beq
\vv{j} = \mu\eta_0 \pder{B_z}{t}\ort{x} -\left(\mu\pder{B_z}{t}+\pder{B_z}{x} \right)\ort{y}\,,
\eeq 
where the same Cartesian basis is used as in section \ref{sec:AW-UPF}. For a monochromatic wave with $B_z=B_{k}e^{i(-\omega t +kx)}$, the corresponding amplitude of the current is 
\beq
j_k = \omega\eta_0 B_k\,.
\eeq
The maximum electric current that can be sustained by the cold electron-positron plasma in the wave is
\beq
j_{max} = n_0\gamma e \,,
\eeq 
and the condition $j_k<j_{max}$ reads exactly the same as the charge starvation condition \eqref{eq:cond1}. 

\section*{Data Availability}     

There are no primary datasets associated with this paper, as all theoretical derivations and calculations are presented directly within the article.

\section*{Acknowledgments}
The author would like to thank Charles Gammie  for sharing his views on the \A wave properties, which was one of the motivations for writing this paper.   
     

\bibliographystyle{mnras}
\bibliography{../BibFiles/plasma,../BibFiles/komissarov,../BibFiles/numerics,../BibFiles/AWaves,../BibFiles/astro,../BibFiles/physics,../BibFiles/hea,../BibFiles/jets,../BibFiles/nstars}

\begin{thebibliography}{34}
\expandafter\ifx\csname natexlab\endcsname\relax\def\natexlab#1{#1}\fi

\bibitem[{Anile}(1989)]{Anile89}
{Anile} A.~M., 1989, {Relativistic fluids and magneto-fluids : with
  applications in astrophysics and plasma physics}, Cambridge ; New York :
  Cambridge University Press

\bibitem[Anile \& Pennisi(1987)]{AnilePenn87}
Anile A.~M., Pennisi S., 1987, Annales de l'I.H.P. Physique théorique, 46, 1,
  27

\bibitem[{Barnes}(1966)]{Barnes66}
{Barnes} A., 1966, Physics of Fluids, 9, 8, 1483

\bibitem[{Belcher} \& {Olbert}(1975)]{BO-75}
{Belcher} J.~W., {Olbert} S., 1975, \apj, 200, 369

\bibitem[{Beloborodov}(2024)]{Beloborodov-24}
{Beloborodov} A.~M., 2024, \apj, 975, 2, 223

\bibitem[{Blaes} et~al.(1989){Blaes}, {Blandford}, {Goldreich} \&
  {Madau}]{BBGM-89}
{Blaes} O., {Blandford} R., {Goldreich} P., {Madau} P., 1989, \apj, 343, 839

\bibitem[{Bransgrove} et~al.(2020){Bransgrove}, {Beloborodov} \&
  {Levin}]{BBL-20}
{Bransgrove} A., {Beloborodov} A.~M., {Levin} Y., 2020, \apj, 897, 2, 173

\bibitem[{Bruno} \& {Carbone}(2013)]{BC13review}
{Bruno} R., {Carbone} V., 2013, Living Reviews in Solar Physics, 10, 1, 2

\bibitem[{Cohen} \& {Kulsrud}(1974)]{CH74}
{Cohen} R.~H., {Kulsrud} R.~M., 1974, Physics of Fluids, 17, 2215

\bibitem[De~Hoffmann \& Teller(1950)]{HT50}
De~Hoffmann F., Teller E., 1950, Phys. Rev., 80, 692

\bibitem[{Del Zanna} et~al.(2007){Del Zanna}, {Zanotti}, {Bucciantini} \&
  {Londrillo}]{DZ07}
{Del Zanna} L., {Zanotti} O., {Bucciantini} N., {Londrillo} P., 2007, \aap,
  473, 1, 11

\bibitem[{Derby}(1978)]{Derby78}
{Derby} Jr. N.~F., 1978, \apj, 224, 1013

\bibitem[{Duncan} \& {Thompson}(1992)]{DT-92}
{Duncan} R.~C., {Thompson} C., 1992, \apjl, 392, L9

\bibitem[{Friedrichs}(1974)]{Fr74}
{Friedrichs} K.~O., 1974, Communications in Pure Applied Mathematics, 27, 749

\bibitem[{Galeev} \& {Oraevskii}(1962)]{GO62}
{Galeev} A., {Oraevskii} V., 1962, Dokl. Akad. Nauk SSSR, 147, 1, 71

\bibitem[{Gazis} et~al.(1994){Gazis}, {Barnes}, {Mihalov} \& {Lazarus}]{GBML94}
{Gazis} P.~R., {Barnes} A., {Mihalov} J.~D., {Lazarus} A.~J., 1994, \jgr, 99,
  A4, 6561

\bibitem[{Goldstein} et~al.(1995){Goldstein}, {Smith}, {Balogh}, {Horbury},
  {Goldstein} \& {Roberts}]{GSB95}
{Goldstein} B.~E., {Smith} E.~J., {Balogh} A., {Horbury} T.~S., {Goldstein}
  M.~L., {Roberts} D.~A., 1995, \grl, 22, 23, 3393

\bibitem[{Goldstein}(1978)]{Gold78}
{Goldstein} M.~L., 1978, \apj, 219, 700

\bibitem[{Holzer} et~al.(1983){Holzer}, {Fla} \& {Leer}]{HFL-83}
{Holzer} T.~E., {Fla} T., {Leer} E., 1983, \apj, 275, 808

\bibitem[{Ishizaki} \& {Ioka}(2024)]{II24}
{Ishizaki} W., {Ioka} K., 2024, \pre, 110, 1, 015205

\bibitem[{Jeffrey} \& {Taniuti}(1964)]{JT64}
{Jeffrey} A., {Taniuti} T., 1964, {Non-linear wave propagation}, New York:
  Academic Press

\bibitem[{Kellogg}(2020)]{Kellogg20}
{Kellogg} P.~J., 2020, \apj, 891, 1, 51

\bibitem[{Komissarov}(1997)]{ssk-aw}
{Komissarov} S.~S., 1997, Physics Letters A, 232, 435

\bibitem[{Komissarov}(1999)]{ssk-godun99}
{Komissarov} S.~S., 1999, \mnras, 303, 343

\bibitem[{Komissarov}(2002)]{ssk-ffde}
{Komissarov} S.~S., 2002, \mnras, 336, 759

\bibitem[{Lax}(1957)]{Lax57}
{Lax} P.~D., 1957, Comm. Pure \& Appl. Math., 10, 537

\bibitem[{Lyutikov}(2021)]{LyutikovCE-21}
{Lyutikov} M., 2021, \apj, 922, 2, 166

\bibitem[{Marriott} \& {Tenerani}(2024)]{MT24}
{Marriott} M., {Tenerani} A., 2024, \apj, 975, 2, 232

\bibitem[{Montgomery} \& {Tidman}(1964)]{MT64}
{Montgomery} D.~C., {Tidman} D.~A., 1964, {Plasma kinetic theory}, New York:
  McGraw-Hill

\bibitem[{Sagdeev} \& {Galeev}(1969)]{SG69}
{Sagdeev} R.~Z., {Galeev} A.~A., 1969, {Nonlinear Plasma Theory}, New York:
  W.A.Benjamin, Inc.

\bibitem[{Thompson} \& {Blaes}(1998)]{TB-98}
{Thompson} C., {Blaes} O., 1998, \prd, 57, 6, 3219

\bibitem[{Whitham}(1974)]{Whitham}
{Whitham} G., 1974, {Linear and Nonlinear waves}, New York: John Wiley \& Sons

\bibitem[{Yasuda} et~al.(2019){Yasuda}, {Suzuki} \& {Kozasa}]{YSK-19}
{Yasuda} Y., {Suzuki} T.~K., {Kozasa} T., 2019, \apj, 879, 2, 77

\bibitem[{Zhang}(2023)]{Zhang23}
{Zhang} B., 2023, Reviews of Modern Physics, 95, 3, 035005

\end{thebibliography}
\end{document}